\documentclass[a4paper]{article}

\usepackage{INTERSPEECH2020}
\usepackage{url}
\usepackage[colorlinks,linkcolor=black,anchorcolor=black,citecolor=black,urlcolor=black]{hyperref}
\usepackage{multirow}

\title{Knowledge-and-Data-Driven Amplitude Spectrum Prediction for Hierarchical Neural Vocoders}
\name{Yang Ai, Zhen-Hua Ling\thanks{This work was partially funded by the National Natural Science
Foundation of China under Grant 61871358.}}
\address{National Engineering Laboratory for Speech and Language Information Processing
\\University of Science and Technology of China, Hefei, P.R.China}
\email{ay8067@mail.ustc.edu.cn, zhling@ustc.edu.cn}

\begin{document}

\maketitle
\begin{abstract}
In our previous work, we have proposed a neural vocoder called HiNet which recovers speech waveforms by predicting amplitude and phase spectra hierarchically from input acoustic features.
In HiNet, the amplitude spectrum predictor (ASP) predicts log amplitude spectra (LAS) from input acoustic features.
This paper proposes a novel knowledge-and-data-driven ASP (KDD-ASP) to improve the conventional one.
First, acoustic features (i.e., F0 and mel-cepstra) pass through a knowledge-driven LAS recovery module to obtain approximate LAS (ALAS).
This module is designed based on the combination of STFT and source-filter theory, in which the source part and the filter part are designed based on input F0 and mel-cepstra, respectively.
Then, the recovered ALAS are processed by a data-driven LAS refinement module which consists of multiple trainable convolutional layers to get the final LAS.
Experimental results show that the HiNet vocoder using KDD-ASP can achieve higher quality of synthetic speech than that using conventional ASP and the WaveRNN vocoder on a text-to-speech (TTS) task.

\end{abstract}
\noindent\textbf{Index Terms}: neural vocoder, log amplitude spectrum, source-filter, TTS

\section{Introduction}

%Speech synthesis, a technology that converts texts into speech waveforms, plays a more and more important role in people's daily life.
Nowadays, statistical parametric speech synthesis (SPSS) has become a popular text-to-speech (TTS) approach thanks to its flexibility and high quality. % achieved by acoustic modeling and vocoder-based waveform generation.
Both acoustic models which predict acoustic features (e.g., mel-cepstra and F0) from texts and vocoders \cite{dudley1939vocoder} which reconstruct speech waveforms from predicted acoustic features are essential in SPSS.
%The performance of vocoders affects the quality of synthetic speech significantly.
Early SPSS systems preferred to adopt conventional vocoders, such as STRAIGHT \cite{kawahara1999restructuring} and WORLD \cite{morise2016world} as their vocoders.
These vocoders are designed based on the source-filter model of speech production \cite{gunnar1960acoustic} and have some limitations, such as the loss of phase information and spectral details.
%Some conventional vocoders, such as STRAIGHT \cite{kawahara1999restructuring} and WORLD \cite{morise2016world} which are designed based on the source-filter model of speech production \cite{gunnar1960acoustic}, have been popularly applied in current SPSS systems.
%However, these vocoders still have some deficiencies, such as the loss of spectral details and phase information.

Recently, some autoregressive neural generative models such as WaveNet \cite{oord2016wavenet}, SampleRNN \cite{mehri2016samplernn} and WaveRNN \cite{kalchbrenner2018efficient} have been proposed
and achieved good performance on generating raw audio signals.
%However, these models always generated waveforms with low quantization bits (e.g., 8-bit or 10-bit) which suffered from perceptible quantization errors.
%In order to achieve 16-bit quantization of speech waveforms, the WaveRNN model \cite{kalchbrenner2018efficient} was proposed, which generated 16-bit waveforms by splitting the recurrent neural network (RNN) state into two parts and predicting the 8 coarse bits and the 8 fine bits respectively.
Their variants such as knowledge-distilling-based models (e.g., parallel WaveNet \cite{oord2017parallel} and ClariNet \cite{ping2018clarinet}) and flow-based models (e.g., WaveGlow \cite{prenger2018waveglow}) were also proposed to further improve the performance and generation efficiency.
Based on these waveform generation models, neural vocoders   
have been developed \cite{tamamori2017speaker,hayashi2017investigation,adiga2018use,ai2018samplernn,ai2019dnn,lorenzo2018robust}, which reconstruct speech waveforms from various acoustic features for SPSS, voice conversion \cite{liu2018wavenet,kobayashi2017statistical}, bandwidth extension \cite{ling2018waveform}, etc. % and speech coding \cite{klejsa2018high}.
Although these neural vocoders outperformed the conventional ones significantly, they still have some limitations. % with current neural vocoders.
The autoregressive neural vocoders have low generation efficiency due to their point-by-point generation process.
For knowledge-distilling-based vocoders and flow-based vocoders, it is difficult to train them due to their complicated training process and high complexity of model structures respectively.

Subsequently, some improved neural vocoders, such as glottal neural vocoder \cite{cui2018new,juvela2018speaker}, LPCNet \cite{valin2019lpcnet}, and neural source-filter (NSF) vocoder \cite{wang2019neural,wang2019neural3,wang2019neural2,zhao2019transferring}, have been further proposed.
These vocoders combine speech production mechanisms with neural networks and have also demonstrated impressive performance.
In our previous work \cite{ai2020neural}, we proposed a neural vocoder named HiNet, which consists of an amplitude spectrum predictor (ASP) and a phase spectrum predictor (PSP).
HiNet produces speech waveforms by first predicting amplitude spectra from input acoustic features using ASP and then predicting phase spectra from amplitude spectra using PSP.
%which recovers speech waveforms by predicting amplitude and phase spectra hierarchically from input acoustic features.
%The HiNet vocoder is composed of an amplitude spectrum predictor (ASP) and a phase spectrum predictor (PSP).
The outputs of ASP and PSP are combined to recover speech waveforms by short-time Fourier synthesis (STFS).
Besides, generative adversarial networks (GANs) \cite{goodfellow2014generative} are also introduced into ASP and PSP to further improve their performance. % of the HiNet vocoder.
Experimental results show that the proposed HiNet vocoder can generate waveforms with high quality and high efficiency.
%However, our previous experiments focused on the analysis-synthesis (AS) task (i.e., using natural acoustic features as input), and gave less research on the text-to-speech (TTS) task (i.e., using predicted acoustic feature as input).

In this paper, we propose a novel knowledge-and-data-driven ASP (KDD-ASP) to replace the conventional one in a HiNet vocoder.
The aim of KDD-ASP is to integrate speech production and analysis knowledge into data-driven LAS prediction, %close the gap between the input ALAS and the target natural LAS of data-driven LAS refinement module,
expecting to improve the accuracy and generalization ability of ASP, especially when predicted acoustic features are used as input. %and promotion ability of the model especially for TTS task.
KDD-ASP consists of a knowledge-driven LAS recovery module and a data-driven LAS refinement module.
The first module is designed based on the combination of STFT and the source-filter theory of speech production, and generates approximate LAS (ALAS) from input acoustic features (i.e., F0 and mel-cepstra).
We assume that the speech signal is produced via a source-filter process \cite{gunnar1960acoustic}.
The source excitation signal and the filter are designed according to the input F0 and mel-cepstra respectively.
Then, ALAS can be calculated by imitating the process of STFT which includes truncation, windowing and FFT.
All operations are performed in the frequency domain.
The second module predicts the final LAS from ALAS.
This module consists of multiple trainable convolutional layers and is trained in a data-driven way.
Experimental results confirm that the HiNet vocoder using KDD-ASP can achieve higher quality of synthetic speech than that using conventional ASP and the WaveRNN vocoder on a TTS task.

This paper is organized as follows.
In Section \ref{sec: HiNet vocoder}, we briefly review the HiNet vocoder \cite{ai2020neural}.
In Section \ref{sec: Knowledge-driven ASP}, we describe the details of our proposed KDD-ASP.
Section \ref{sec: Experiments} reports our experimental results.
Conclusions are given in Section \ref{sec: Conclusion}.

\section{HiNet vocoder}
\label{sec: HiNet vocoder}

HiNet \cite{ai2020neural} is a novel neural vocoder which recovers speech waveforms by predicting amplitude and phase spectra hierarchically from input acoustic features.
Conventional neural vocoders usually employ single neural networks to generate speech waveforms directly.
In contrast, the HiNet vocoder consists of an amplitude spectrum predictor (ASP) and a phase spectrum predictor (PSP).
%Different from existing neural vocoders which directly generate waveform samples using single neural networks, the HiNet vocoder is composed of an amplitude spectrum predictor (ASP) and a phase spectrum predictor (PSP).
ASP uses acoustic features as input and predicts frame-level log amplitude spectra (LAS). Then PSP uses the predicted LAS and F0 as input and recovers the phase spectra.
%ASP predicts frame-level log amplitude spectra (LAS) from acoustic features. Then, the predicted LAS and F0 are sent into the PSP for phase recovery.
Finally, the outputs of ASP and PSP are combined to recover speech waveforms by short-time Fourier synthesis (STFS).

In our implement, ASP is a simple non-autoregressive DNN containing multiple feed-forward (FF) layers.
%In our implement, ASP is a simple DNN consisting of multiple feed-forward (FF) layers without any autoregressive structures.
It concatenates the acoustic features at current and previous frames as input to predict the LAS at current frame.
At the training stage, the target LAS are extracted from natural waveforms by STFT.
A GAN criterion is adopted to build ASP.
The DNN model is used as the generator of GAN and its discriminator consists of multiple convolutional layers which operate along the frequency axis of the input LAS.
A Wasserstein GAN \cite{gulrajani2017improved} loss is combined with the mean square error (MSE) between the predicted LAS and natural ones to train the generator. % is also used as an auxiliary loss in the G loss.
%The training process is divided into three steps, first using the MSE loss to train the generator, then using the classification loss to train the discriminator, and finally using complete losses to train the generator and discriminator alternately.

PSP is constructed by concatenating a neural waveform generator with a phase spectrum extractor.
The neural waveform generator is built by adapting the NSF vocoder \cite{wang2019neural} from three aspects, 1) using LAS as the input, 2) pre-calculating the initial phase of the sine-based excitation signal for each voiced segment at the training stage and 3) adopting a combined loss function including MSE on amplitude spectra, waveform loss and correlation loss.
GAN is also introduced into PSP.
Here, the neural waveform generator of PSP is used as the generator of GAN and its discriminator is similar with that of ASP except that its input features are waveforms instead of LAS.
%A Wasserstein GAN \cite{gulrajani2017improved} loss is used for training and the combined loss is also used as an auxiliary loss in G loss.
%The training process is using G and D loss to train the generator and discriminator alternately.
%
%\subsection{Structural improvements for ASP}
%\label{subsec: Structural improvements for ASP}
%
%After this work \cite{ai2020neural}, we made further structural improvements to ASP.
%The improved ASP's structure is shown in Fig. \ref{fig: ASP}.
%There are two main improvements here.
%\begin{itemize}
%\item For the generator, convolutional layers are used instead of FF layers. The input is the acoustic features at current frame instead of the concatenated ones.
%\item Add another discriminator which operate along with the time axis of the input LAS.
%\end{itemize}
%We have confirmed that these improvements are effective and we adopted the improved version of ASP in all the experiments in this paper.

\section{Knowledge-and-Data-Driven ASP}
\label{sec: Knowledge-driven ASP}
\begin{figure}[t]
    \centering
    \includegraphics[height=2.1cm]{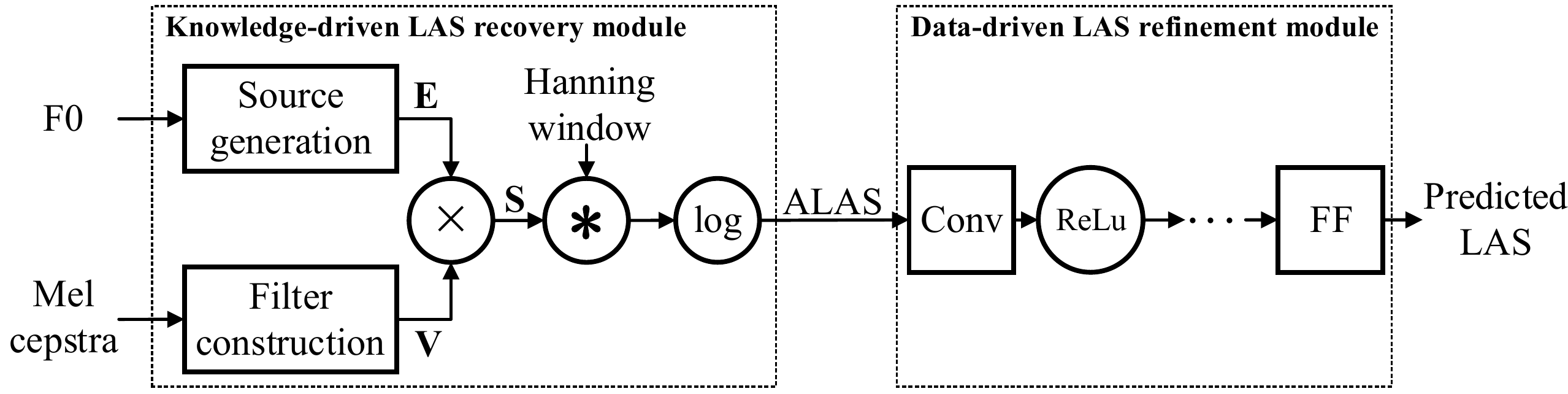}
    \caption{Model structure of KDD-ASP. Here, $\times$, $*$ and $log$ denote element-wise product, convolutional and log operation respectively, \emph{FF} and \emph{Conv} represent feed-forward and convolutional layers respectively and \emph{ReLu} means rectified linear units.}
    \label{fig: KD-ASP}
\end{figure}

This paper proposes a novel knowledge-and-data-driven ASP (KDD-ASP) to replace the conventional one in a HiNet vocoder.
The KDD-ASP is constructed by concatenating a knowledge-driven LAS recovery module with a data-driven LAS refinement module as shown in Fig. \ref{fig: KD-ASP}.

\subsection{Knowledge-driven LAS recovery module}
\label{subsec: Knowledge-driven LAS recovery module}

The equation for extracting LAS directly from a signal $\bm{s}$ by STFT can be written as follows,
\begin{align}
\label{equ: LAS}
\bm{LAS}_n=\log|\mathcal{F}(\bm{s}_n\odot \bm{w})|,
\end{align}
where $\bm{s}_n=[s_{n,1},\dots,s_{n,L}]^\top$ and $\bm{LAS}_n=[LAS_{n,1}\dots,LAS_{n,K}]^\top$ are the framed signal of $\bm{s}$ and the LAS at the $n$-th frame respectively, and
$\bm{w}=[w_1,\dots,w_L]^\top$ denotes the Hanning window for short-time analysis. $L$ is the frame number.
$K=\frac{FN}{2}+1$ represents the number of frequency bins and $FN$ is the FFT point number. $\odot$ and $\mathcal{F}$ represent element-wise product and FFT, respectively.

Inspired by this process, the knowledge-driven LAS recovery module constructs approximate LAS (ALAS) from F0 and mel-cepstra based on the frequency-domain representation of Eq. (\ref{equ: LAS}).
We assume that the speech signal at the $n$-th frame $\bm{s}_n$ is obtained by the convolution between a source excitation signal $\bm{e}_n$ and a filter impulse response $\bm{v}_n$.
In frequency domain, this process can be represented  as
\begin{align}
\label{equ: S}
\bm{S}_n=\bm{E}_n\odot \bm{V}_n,
\end{align}
where $\bm{S}_n=[S_{n,1},\dots,S_{n,K}]^\top$, $\bm{E}_n=[E_{n,1},\dots,E_{n,K}]^\top$ and $\bm{V}_n=[V_{n,1},\dots,V_{n,K}]^\top$ are the Fourier transform of $\bm{s}_n$, $\bm{e}_n$ and $\bm{v}_n$ respectively.

\begin{figure*}[t]
    \centering
    \includegraphics[height=3cm]{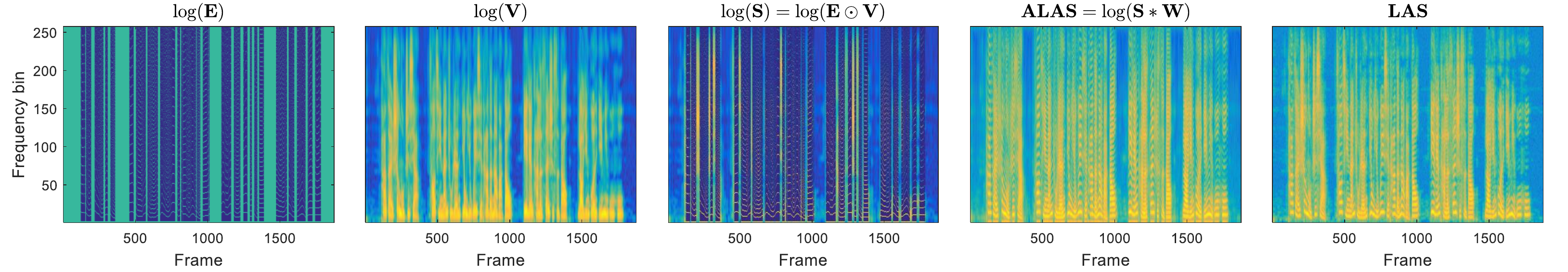}
    \caption{The visualization of $\log(\bm{E}_n)$, $\log(\bm{V}_n)$, $\log(\bm{S}_n)$, $\bm{ALAS}_n$ and $\bm{LAS}_n$ for an example utterance. Here, the input F0 and mel-cepstra are natural ones.}
    \label{fig: Color_map}
\end{figure*}

Let $f_n$ denote the F0 value of the $n$-th frame when it is voiced and $f_n=0$ when the frame is unvoiced.
%$E_n$ is designed according to F0 $f_n$.
For voiced frames ($f_n>0$), $\bm{E}_n$ is produced as a pulse train %(amplitude=1)
with equal frequency interval $K_0=Round(\frac{f_n}{F_s} \cdot FN)$, which corresponds to constructing all the harmonics below the Nyquist frequency, where $F_s$ is the sampling rate.
For unvoiced frames ($f_n=0$), we set $E_n\equiv 1$, meaning that the excitation signal is a Gaussian white noise.
The equation for producing $\bm{E}_n$ based on F0 values can be written as
\begin{align}
\label{equ: E}
E_{n,k}=\left\{\begin{array}{ll}1,&f_n>0, k=i\cdot K_0\\
0,&f_n>0, k\neq i\cdot K_0 \text{ or } f_n=0\end{array}\right.,
\end{align}
where $i=1,2,\dots,\lceil \frac{K}{K_0}\rceil$.

$\bm{V}_n$ is calculated by transforming mel-cepstra to amplitude spectra \cite{tokuda1994mel}.
%The first step is to transform mel-cepstra to cepstra.
The mel-cepstral coefficients at the $n$-th frame (with energy as the first order)
are first padded with zeros to form a $K$-dimensional vector $\bm{m}_n=[m_{n,1},\dots,m_{n,K}]^\top$.
Then, the cepstral coefficients $c_{n,k}, k=1,\dots,K$ are calculated by the following iterative formulas
\begin{align}
\label{equ: cepstra}
c_{n,k}(i)\!=\!\left\{\begin{array}{ll}\!m_{n,i}\!-\!\alpha\!\cdot\! c_{n,1}(i\!+\!1),&k\!=\!1\\
\!(1\!-\!\alpha^2)\!\cdot\! c_{n,1}(i\!+\!1)\!-\!\alpha\!\cdot\! c_{n,2}(i\!+\!1),&k\!=\!2\\
\!c_{n,k\!-\!1}(i\!+\!1)\!-\!\alpha\!\cdot\![c_{n,k}(i\!+\!1)\!-\!c_{n,k\!-\!1}(i)],&k\!>\!2\end{array}\right.\!,
\end{align}
where $i$ iterates from $K$ to 1 with the initial value $c_{n,k}(K+1)=0, k=1,\dots,K$.
$\alpha$ is the mel-frequency warping coefficient, which is 0.42 for $F_s=16000$. %, $\alpha=0.42$.
After the iteration, we can obtain the cepstra vector $\bm{c}_n=[c_{n,1}(1),\dots,c_{n,K}(1)]^\top$, which is further transformed to the amplitude spectra $\bm{V}_n$ by
\begin{align}
\label{equ: V}
\bm{V}_n=\exp[\mathcal{F}(\bm{c}_n)].
\end{align}

Finally, ALAS can be calculated as
\begin{align}
\label{equ: ALAS}
\bm{ALAS}_n=\log|\bm{S}_n*\bm{W}|,
\end{align}
where $\bm{ALAS}_n=[ALAS_{n,1}\dots,ALAS_{n,K}]^\top$ is the $n$-th frame ALAS and $\bm{W}=[W_1,\dots,W_K]^\top$ is the Fourier transform of the analysis window $\bm{w}$. The operation $*$ represents convolution.
It is worth mentioning that the elements in the vectors of $\bm{S}_n$ and $\bm{W}$ should be rearranged by complementing their mirror-symmetric parts and shifting the zero-frequency component to the center before convolution.

%Fig. \ref{fig: Color_map} shows an visualization of $E$, $V$, $S$, $ST=|S*W|$, $ALAS$ and $LAS$ on all frames ($n=1,\dots,N$) for an utterance example.
%We can see that the ALAS is close to LAS, meaning that the input and output of ASP are similar, expecting to make it easier to train ASP.

\subsection{Data-driven LAS refinement module}
\label{subsec: Data-driven LAS refinement module}

The data-driven LAS refinement module converts ALAS to final LAS by a trainable neural network.
In our implement, this module adopts the ASP model in Section \ref{sec: HiNet vocoder} but has two structural improvements.
%\begin{itemize}
First, convolutional layers are used instead of FF layers in the generator and the input is the ALAS at current frame instead of the concatenated ones as shown in Fig. \ref{fig: KD-ASP}.
Second, another discriminator which operates along with the time axis of the input LAS is added\footnote{Discriminators are not shown in Fig. \ref{fig: KD-ASP} for simplification.}.
%\end{itemize}

\section{Experiments}
\label{sec: Experiments}

\subsection{Experimental conditions}
\label{subsec: Experimental conditions}

A Chinese speech synthesis  corpus with 13334 utterances ($\sim$20 hours)  was used in our experiments.
The speaker was a female and the waveforms had 16 kHz sampling rate with 16 bits resolution.
The training, validation and test sets contained 13134, 100 and 100 utterances, respectively. %, and the remaining 100 utterances were used as the test set.
%We select 13134 and 100 utterances to construct the training set and validation set respectively, and the remaining 100 utterances were used as the test set.
The natural acoustic features were extracted with a frame length and shift of 25 ms and 5 ms respectively.
%The natural acoustic features were extracted by STRAIGHT and the window size was 25ms and the window shift was 5ms.
The acoustic features at each frame were 43-dimensional including 40-dimensional mel-cepstra, an energy, an F0 and a V/UV flag.
For SPSS, a bidirectional LSTM-RNN acoustic model \cite{fan2014tts} having 2 hidden layers with 1024 units per layer (512 forward units and 512 backward units) was trained as the acoustic model, which predicted acoustic features from 566-dimensional linguistic features. 
The output of the acoustic model was 127-dimensional including 43-dimensional acoustic features together with their delta and acceleration counterparts (the V/UV flag had no dynamic components).
%The output of the acoustic model contained acoustic features together with their delta and acceleration counterparts, which were totally 127 dimensions (the V/UV flag had no dynamic components).
Then, the predicted acoustic features were generated from the output by maximum likelihood parameter generation (MLPG) \cite{tokuda2000speech} considering global variance (GV) \cite{toda2007speech}.
Since this paper focuses on vocoders, natural durations obtained by HMM-based forced alignment were used at synthesis time.

Three vocoders were compared in our experiments\footnote{Examples of generated speech can be found at \url{http://home.ustc.edu.cn/~ay8067/Interspeech2020/demo.html}.}.
The descriptions of these vocoders are as follows.

%\begin{itemize}
%\item
1) \emph{\textbf{WaveRNN}} A 16-bit WaveRNN-based neural vocoder using acoustic features as input. 
    This vocoder was implemented by ourselves and the efficiency optimization strategies \cite{kalchbrenner2018efficient} were not adopted here.
    Its structure was the same as \emph{\textbf{WaveRNN}} in our previous work \cite{ai2020neural} which performed better than the 16-bit WaveNet vocoder using open source implementation\footnote{\url{https://github.com/r9y9/wavenet_vocoder}.}.
    The waveform samples were quantized to discrete values by 16-bit linear quantization and the model had one hidden layer with 1024 nodes where 512 nodes for coarse outputs and another 512 nodes for fine outputs.
%    The waveform samples were quantized to discrete values by 16-bit linear quantization and the built model had one hidden layer of 1024 nodes where 512 nodes for coarse outputs and another 512 nodes for fine outputs.
    Models were trained and evaluated  on a single Nvidia 1080Ti GPU using TensorFlow \cite{abadi2016tensorflow}.

2) \emph{\textbf{HiNet}} A HiNet vocoder using conventional ASP.
The structure of ASP is the same with that of the data-driven LAS refinement module introduced in Section \ref{subsec: Data-driven LAS refinement module}. %in \emph{\textbf{HiNet-KDD}}.
When extracting natural LAS, the frame length and frame shift of STFT were 20ms (i.e., $L=320$) and 5ms respectively and FFT point number was 512 (i.e., $K=257$).
There were 3 convolutional layers with 2048 nodes per layer (filter width=7), and a 257-dimensional linear output layer which predicted the LAS.
For each training step, ASP used 128 frames of acoustic features as input and outputted corresponding 128 frames of LAS.
GANs were also used in ASP.
Discriminator \#1 operated along with the frequency axis and consisted of 6 convolutional layers (filter width=9, stride size=2) and their channels were 16, 32, 64, 128 and 256 respectively.
The resulting dimensions per layer, being it frequency bins $\times$ channels, were 257$\times$1, 129$\times$16, 65$\times$32, 33$\times$64, 17$\times$128 and 9$\times$256.
Finally, two FF layers with 256 and 9 nodes respectively were used to map the 9$\times$256 convolutional results into a value for loss calculation.
Discriminator \#2 operated along with the time axis and consisted of 4 convolutional layers (filter width=9, stride size=2) and their channels were 64, 128, 256 and 512 respectively.
The resulting dimensions per layer, being it frequency bins $\times$ channels, were 128$\times$257, 64$\times$64, 32$\times$128, 16$\times$256 and 8$\times$512.
Finally, two FF layers with 512 and 8 nodes respectively were used to map the 8$\times$512 convolutional results into a value for loss calculation.
Remaining settings of ASP and all the settings of PSP are the same as the \emph{\textbf{HiNet-S-GAN}} vocoder in our previous work \cite{ai2020neural}.
ASP and PSP models were both trained and evaluated on a single Nvidia 1080Ti GPU using TensorFlow framework \cite{abadi2016tensorflow}.

3) \emph{\textbf{HiNet-KDD}} A HiNet vocoder using the KDD-ASP proposed in this paper.
For KDD-ASP, the knowledge-driven LAS generation module adopted the same settings with that of extracting natural LAS (i.e., $L=320$ and $K=257$) and the settings of the data-driven LAS refinement module
%has been shown in
were the same as the ASP of \emph{\textbf{HiNet}}.
The settings of PSP and the implementation conditions were all the same as that of \emph{\textbf{HiNet}}.
Fig. \ref{fig: Color_map} shows the visualization of $\bm{E}_n$, $\bm{V}_n$, $\bm{S}_n$, $\bm{ALAS}_n$ and $\bm{LAS}_n$ for all frames in an example utterance.
We can see that the recovered ALAS is close to the reference LAS with analogous harmonic and formant structures, meaning that the input and output of the data-driven LAS refinement module are similar,
expecting to facilitate the model learning and to improve the performance of predicting amplitude spectra.

%\end{itemize}

\begin{table}
\centering
    \caption{Objective evaluation results of \emph{\textbf{WaveRNN}}, \emph{\textbf{HiNet}} and \emph{\textbf{HiNet-KDD}} on the test set. ``AS" stands for analysis-synthesis task and ``TTS" stands for TTS task.}
    \resizebox{7.5cm}{1.6cm}{
    \begin{tabular}{c c | c c c}
        \hline
        \hline
         & & \emph{\textbf{WaveRNN}}& \emph{\textbf{HiNet}}& \emph{\textbf{HiNet-KDD}}\\
         \hline
         \multirow{5}{*}{\emph{AS}}
         & SNR(dB) & 4.6631 & \textbf{5.2587} & 5.0152\\
         & LAS-RMSE(dB) & 4.9623 & \textbf{4.2602} & 4.5659\\
         & MCD-V(dB) & 1.0702 & \textbf{0.7686} & 0.8583\\
         & F0-RMSE(cent) & 13.2365 & 9.3345 & \textbf{9.0960}\\
         & V/UV error(\%) & 4.2515 & 2.0116 & \textbf{2.0041}\\
         \cline{1-5}
         \multirow{3}{*}{\emph{TTS}}
         & MCD-V(dB) & 1.0702 & 1.0939 & \textbf{0.9488}\\
         & F0-RMSE(cent) & 12.4645 & 7.0877 & \textbf{6.4970}\\
         & V/UV error(\%) & 3.5247 & \textbf{1.7983} & 2.0194\\
        \hline
        \hline
    \end{tabular}}
\label{tab: objective}
\end{table}

\subsection{Objective evaluation}
\label{subsec: Objective evaluation}

We first compared the performance of these three vocoders using objective evaluations.
Five objective metrics used in our previous work \cite{ai2020neural} were adopted here, including signal-to-noise ratio (SNR), root MSE (RMSE) of LAS (denoted by LAS-RMSE), mel-cepstrum distortion for voiced frames (denoted by MCD-V), MSE of F0 (denoted by F0-RMSE) and V/UV error.
For the analysis-synthesis (AS) task, the references are natural waveforms or the acoustic features extracted from natural waveforms.
For the TTS task, the references are the mel-cepstra and F0 predicted by the acoustic model and only MCD-V, F0-RMSE and V/UV error were adopted since the calculation of SNR and LAS-RMSE relied on natural speech waveforms.

\begin{figure}
    \centering
    \includegraphics[height=5.9cm]{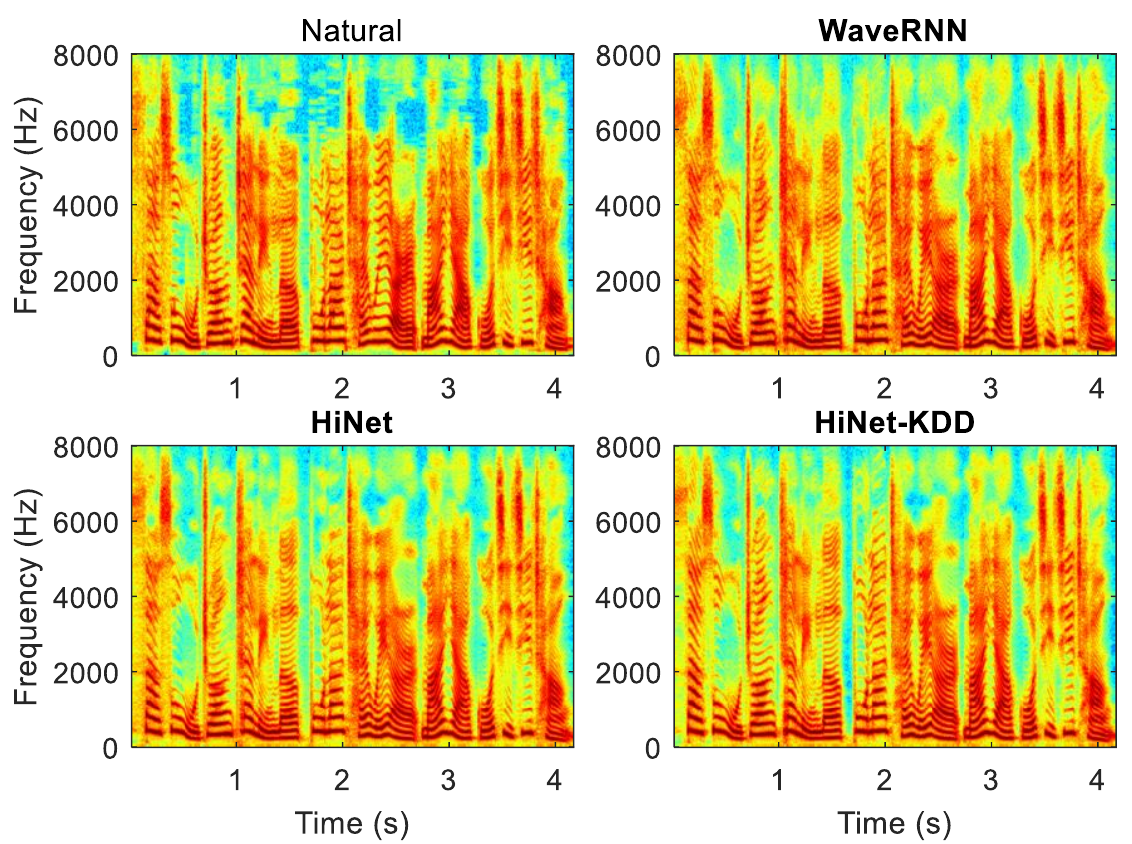}
    \caption{The spectrograms of natural speech and the speech generated by \emph{\textbf{WaveRNN}}, \emph{\textbf{HiNet}} and \emph{\textbf{HiNet-KDD}} on TTS task for an example sentence in the test set.}
    \label{fig: Spectrogram}
\end{figure}

The objective results on the test set are listed in Table \ref{tab: objective}.
It is obvious that both \emph{\textbf{HiNet}} and \emph{\textbf{HiNet-KDD}} outperformed \emph{\textbf{WaveRNN}} on most metrics for both AS and TTS tasks.
By comparing \emph{\textbf{HiNet}} and \emph{\textbf{HiNet-KDD}}, we can find that \emph{\textbf{HiNet-KDD}} performed better on F0-RMSE %had more advantages on F0 recovering (better F0-RMSE)
than \emph{\textbf{HiNet}} for both AS and TTS tasks,
which indicated that \emph{\textbf{HiNet-KDD}} is better at restoring harmonics for voiced frames.
Considering the SNR, LAS-RMSE and MCD-V for AS task, \emph{\textbf{HiNet-KDD}} was not as good as \emph{\textbf{HiNet}}.
However, for TTS task, \emph{\textbf{HiNet-KDD}} achieved better MCD-V than \emph{\textbf{HiNet}}.
This advantage can be attributed to that using ALAS as the input to train the ASP model improves its generalization ability when dealing with unseen acoustic features.
We also draw the spectrograms extracted from natural waveforms and from the waveforms generated by these three vocoders on TTS task in Fig. \ref{fig: Spectrogram}.
We can see that \emph{\textbf{HiNet-KDD}} can restore more clear harmonics (e.g., 0.7$\sim$1.0s and 1.7$\sim$2.0s)  especially in the high-frequency band than the other two vocoders.

%For analysis-synthesis task, it is obvious that \emph{\textbf{WaveRNN}} achieved the worst performance on all metrics.
%Our proposed \emph{\textbf{HiNet-KD}} outperformed \emph{\textbf{HiNet}} on the F0-RMSE and V/UV error metrics, but the conclusion was the opposite on SNR, LAS-RMSE and MCD.
%This indicated that \emph{\textbf{HiNet-KD}} had more advantages on F0 recovering than \emph{\textbf{HiNet}}.
%Although \emph{\textbf{HiNet-KD}} had worst performance on MCD, it showed the best results on MCD-V.
%This indicated that \emph{\textbf{HiNet-KD}} had more advantages on restoring harmonics for voiced frames than other two vocoders.
%This advantage can be attributed to the ALAS with explicit harmonic components.

\subsection{Subjective evaluation}
\label{subsec: Subjective evaluation}

\begin{table}
\centering
    \caption{Average preference scores (\%) on naturalness among different vocoders, where N/P stands for ``no preference" and $p$ denotes the $p$-value of a $t$-test between two vocoders. ``AS" stands for analysis-synthesis task and ``TTS" stands for TTS task.}
    \resizebox{7.5cm}{1.15cm}{
    \begin{tabular}{c |c c c c c}
        \hline
        \hline
        & \emph{\textbf{WaveRNN}}& \emph{\textbf{HiNet}}& \emph{\textbf{HiNet-KDD}}
         &\emph{N/P}& $p$ \\
         \hline
         \multirow{2}{*}{\emph{AS}}
         & 2.73 & \textbf{72.73} & -- & 24.54 & $<0.01$\\
         & -- & 21.36 & 15.45 & 63.19 & 0.15\\
         \hline
         \multirow{3}{*}{\emph{TTS}}
         & 16.82 & \textbf{57.27} & -- & 25.91 & $<0.01$\\
         & 10.91 & -- & \textbf{66.82} & 22.27 & $<0.01$\\
         & -- & 14.55 & \textbf{53.64} & 31.81 & $<0.01$\\
         \hline
        \hline
    \end{tabular}}
\label{tab: ABX}
\end{table}

Five groups of ABX preference tests were conducted to compare the subjective performance of different vocoders.
In each subjective test, 20 utterances generated by two comparative vocoders were randomly selected from the test set.
Each pair of generated speech were evaluated in random order.
11 Chinese native speakers were asked to judge which utterance in each pair had better naturalness or there was no preference.
The $p$-value of a $t$-test was also calculated to measure the significance of the difference between two comparative vocoders.
%In addition to calculating the average preference scores, the $p$-value of a $t$-test was used to measure the significance of the difference between two vocoders.

The subjective results are shown in Table \ref{tab: ABX}.
We can see that \emph{\textbf{HiNet}} outperformed \emph{\textbf{WaveRNN}} very significantly ($p <$ 0.01) on both AS and TTS tasks.
However, the preference difference between these two vocoders became weaker on TTS task than on AS task.
Comparing \emph{\textbf{HiNet}} with \emph{\textbf{HiNet-KDD}}, we can see that there was no significant difference ($p >$ 0.05) between these two vocoders on AS task but  \emph{\textbf{HiNet-KDD}} outperformed \emph{\textbf{HiNet}} significantly ($p <$ 0.01) on TTS task.
We also conducted a group of ABX test between \emph{\textbf{WaveRNN}} and \emph{\textbf{HiNet-KDD}} for TTS task and \emph{\textbf{HiNet-KDD}} also outperformed \emph{\textbf{HiNet}} significantly ($p <$ 0.01).
Besides, the preference score difference between \emph{\textbf{HiNet-KDD}} and \emph{\textbf{WaveRNN}} was larger than that between \emph{\textbf{HiNet}} and \emph{\textbf{WaveRNN}}.
These results all indicated that using KDD-ASP in HiNet vocoder was  helpful for improving the quality of reconstructed speech waveforms %compared with using the conventional ASP
when the input acoustic features were predicted for TTS.
%For AS task, we first compared the performance between \emph{\textbf{HiNet}} and \emph{\textbf{HiNet-KD}}.
%There was no significant difference ($p >$ 0.05) between the subjective quality of these two vocoders, thus we then only compared \emph{\textbf{HiNet}} with \emph{\textbf{WaveRNN}} and \emph{\textbf{HiNet}} outperformed \emph{\textbf{WaveRNN}} very significantly ($p <$ 0.01). This is consistent with the objective results in Table \ref{tab: objective}.
%For TTS task, \emph{\textbf{HiNet-KD}} outperformed \emph{\textbf{HiNet}} very significantly ($p <$ 0.01), which indicated that using ALAS as ASP's input rather than acoustic features was very helpful for improving the quality of synthetic speech for TTS task.
%And both \emph{\textbf{HiNet}} and \emph{\textbf{HiNet-KD}} outperformed \emph{\textbf{WaveRNN}} very significantly ($p <$ 0.01), thus the HiNet vocoder had better performance than that of the WaveRNN vocoder.
%But the difference between \emph{\textbf{HiNet-KD}} and \emph{\textbf{WaveRNN}} was more significant, which also indicated that using ALAS can indeed further improve the performance of the HiNet vocoder for TTS task.

\section{Conclusion}
\label{sec: Conclusion}

In this paper, we have proposed a novel knowledge-and-data-driven amplitude spectrum predictor (KDD-ASP) to replace the conventional one in HiNet, a hierarchical neural vocoder.
KDD-ASP consists of a knowledge-driven LAS recovery module and a data-driven LAS refinement module.
The first module is designed based on the combination of STFT and source-filter theories in order to convert F0 and mel-cepstra into approximate log amplitude spectra (ALAS).
The input F0 values are used to produce the source signal and the filter part is calculated from mel-cepstra.
%All operations are performed in the frequency domain.
The second module is a convolutional neural network which adopts GANs and predicts the final LAS from input ALAS.
Experimental results show that %the advantages of KDD-ASP were mainly reflected in TTS task and
the HiNet vocoder using KDD-ASP can achieve higher quality of synthetic speech than the HiNet vocoder using conventional ASP and the WaveRNN vocoder on a TTS task.
To explore other knowledge-driven methods for ASP and further improve the performance of phase spectrum prediction will be the tasks of our future research.

\bibliographystyle{IEEEtran}

\bibliography{mybib}

\end{document}